\documentclass[preprint]{aastex}

\shorttitle{The discovery of new warm debris disks around F-type stars.}

\shortauthors{Mo\'or et al.}

\begin{document}


\title{The discovery of new warm debris disks around F-type stars.}

\author{A. Mo\'or\altaffilmark{1}}
\email{moor@konkoly.hu}
\author{D. Apai\altaffilmark{2}}
\author{I. Pascucci\altaffilmark{3}}
\author{P. \'Abrah\'am\altaffilmark{1}}
\author{C. Grady\altaffilmark{4,5}} 
\author{Th.~Henning\altaffilmark{6}}
\author{A. Juh\'asz\altaffilmark{6}} 
\author{Cs. Kiss\altaffilmark{1}}
\author{\'A. K\'osp\'al\altaffilmark{7}}
\altaffiltext{1}{Konkoly Observatory of the Hungarian Academy of Sciences, P.O. Box 67, H-1525 Budapest, Hungary}
\altaffiltext{2}{Space Telescope Science Institute, 3700 San Martin Drive, Baltimore, MD 21218, USA}
\altaffiltext{3}{Department of Physics and Astronomy, Johns Hopkins University, Baltimore, MD21218, USA}
\altaffiltext{4}{NASA Goddard Space Flight Center, Code 667, Greenbelt, MD 20771, USA}
\altaffiltext{5}{Eureka Scientific, 2452 Delmer Street, Suite 100, Oakland, CA 94602, USA}
\altaffiltext{6}{Max-Planck-Institut f\"ur Astronomie, K\"onigstuhl 17, 69117 Heidelberg, Germany}
\altaffiltext{7}{Leiden Observatory, Leiden University, Niels Bohrweg 2, NL-2333 CA Leiden, The Netherlands}


\begin{abstract}
We report the discovery of four rare debris disks with warm excesses around F-stars, 
significantly increasing the number of such systems known in the solar neighborhood.
Three of the disks are consistent with the predictions of steady state planetesimal disk evolution models.
The oldest source, HD\,169666, displays a dust fractional luminosity too high to be in steady state and we 
suggest that this system recently underwent a transient event of dust production.
In addition, two spectra of this star separated by $\sim$three years show silicate emission features, 
indicative of submicron- to micron-sized grains. 
We argue that such small grains would be rapidly depleted and their presence in both spectra suggests 
that the production of small dust is continuous over at least on few years timescale.
We predict that systems showing variable mid-infrared spectra, if exist, will provide valuable help in distinguishing 
the possible scenarios proposed for dust replenishment.

\end{abstract}





\section{Introduction} \label{intro}

Many main-sequence stars exhibit excess emission at mid-to-far infrared (IR) wavelengths, 
implying the existence of a ring or disk of circumstellar dust grains, which re-radiate the 
absorbed starlight at longer wavelengths. Due to mutual collisions, 
radiation pressure and the Poynting-Robertson (PR) drag, the lifetime of the 
emitting dust grains is much shorter than the age of the star, and 
the dust particles are thought to be continuously replenished 
by the erosion of larger planetesimals 
\citep{backman93,wyatt08rev}.
Thus, the existence of a dust 'debris' disk 
implies the presence of a planetesimal belt around the star.   
   
In most known debris
disks around F, G, and K-type stars the dust grains are relatively cold (50--90\,K), 
similar in temperature to the dust located in the Kuiper Belt in our solar system \citep{carpenter08}.
Interestingly, warm debris disks -- where the emission (in $\rm F_{\nu}$) peaks at $\lambda>40$\,$\mu$m, 
corresponding to $T_{dust}\gtrsim 125$\,K -- are rare.
If the infrared emission in these warm disks originated from blackbody grains, 
they would be located at a radius of 1-10\,AU.
Thus, in 'solar-system terminology' these disks are either similar to the main asteroid belt
or are situated between the main belt and the Kuiper Belt. 

So far $\sim$20 warm debris disks have been discovered around FGK-type stars. 
Most of them encircle young stars, that are often members of known young open clusters or 
moving groups  typically located at larger ($>$100\,pc) distances 
\citep{chen05,currie07,currie08,gorlova07,hines06,lisse08,rhee08,smith08}. 
In these young ($<$100\,Myr) systems the source of warm debris dust in the region of terrestrial planets may be an 
outwards propagating ring of 
planetesimal formation and/or the products of massive protoplanet collisions \citep{kb04,meyer08}. 
A smaller number of warm disks have been identified around older solar-type stars 
\citep{song05,beichman05,beichman06,wyatt05}.
All these old systems exhibit high fractional luminosity ($\frac{L_{IR}}{L_{bol}} > 10^{-4}$), 
that cannot be explained 
with a steady-state asteroid belt evolution model, where the planetesimals are
co-located with the warm dust. Instead, \citet{wyatt07} suggested 
that these systems are in a transient state. Recent collisions between large asteroids or the erosion/sublimation of 
planetesimals scattered from an outer reservoir into the inner regions due to a dynamical instability 
are proposed as the origin of the transient warm dust \citep{song05,wyatt07}.
One can find possible analogous transient events also in the history of our solar system \citep{gomes05,
nesvorny03}.

In our extensive program studying debris disks around 78 F-type stars (PID: 3401, 20707, Mo\'or et al., in prep., 2009)
we identified four systems harboring warm disks. All four are new discoveries and are located relatively close 
to the Sun ($<$60\,pc).
Here we analyze their properties, discuss their possible origins and speculate about 
the implications on the formation and evolution of dust in the inner region of planetary systems. 

\section{Observations and data reduction} \label{obsanddatared}

The four stars, \object[HD 13246]{HD\,13246}, \object[HD 53842]{HD\,53842}, 
\object[HD 152598]{HD\,152598}, and \object[HD 169666]{HD\,169666} were selected for our program 
because of earlier hints for IR excess from IRAS and/or their membership in young 
moving groups. \object[HD 13246]{HD\,13246} and \object[HD 53842]{HD\,53842} belong to the Tucana-Horologium 
moving group \citep{zucksong04} implying an age of $\sim$30\,Myr. Age indicators, such as lithium abundances 
and X-ray luminosities, further support this age.  
The galactic motion of \object[HD 152598]{HD\,152598} (U\,=\,$-$10.7, V\,=\,$-$22.6, W\,=\,$-$4.0\,kms$^{-1}$) 
as well as its galactic position are consistent with those of the "b3" subgroup of the Local Association, implying an age of 0.21$\pm$0.07\,Gyr \citep{asiain99}. 
Both its fractional X-ray luminosity ({\sl ROSAT} J165258.2+314203, $\log(L_x/L_{bol})\sim -5.63$) and 
the equivalent width of the lithium 6708\,\AA\, feature 
\citep[43$\pm$8\,m\AA\, from spectra available in the Elodie archive, ][]{moultaka04}
support {this} age.
In the case of \object[HD 169666]{HD\,169666} isochrone fitting yields $\rm 2.1^{+0.2}_{-0.4}$\,Gyr 
\citep{holmberg07}. Its location in the HRD indicates that this star is already leaving the main-sequence, 
supporting the conclusion of the isochrone dating. \object[HD 169666]{HD\,169666} coincides with {\sl ROSAT} J181909.0+713054, its 
fractional X-ray luminosity is $\log(L_x/L_{bol})\sim -5.67$. We note that this value is not inconsistent with 
the measured $\log(L_x/L_{bol})$ values of stars with similar $B-V$ color indices 
in the 1.9\,Gyr old NGC\,752 open cluster \citep{daniel94,giardino08}, 
giving support to the isochrone age determination.
The key properties of the four stars are listed in Table~\ref{stellarprop}.

\subsection{Spitzer observations and data reduction}

Mid- and far-IR images were obtained using the MIPS instrument \citep{rieke04} in small-field 
photometry mode. All four objects were measured at 24 and 70\,$\mu$m, while two of them were also
observed at 160\,$\mu$m.
In addition, low-resolution spectra (R = 70--120) were obtained for each source using the Infrared Spectrograph \citep[IRS,][]{houck04}.
All these observations belong to our program PID\,20707 except the IRS spectrum 
of \object[HD 13246]{HD\,13246}, which was taken from the {\sl Spitzer} Archive (PID: 241, PI: C. Chen). 
 In order to improve the signal-to-noise ratio and search for possible variability 
we performed follow-up IRS measurements for \object[HD 13246]{HD\,13246} and 
\object[HD 169666]{HD\,169666}, obtained 41 and 35 months after the first epochs, respectively
(PID: 50538, PI: \'A. K\'osp\'al).

\paragraph{MIPS:} 
We started the data processing with the BCD files (pipeline version S16.1). 
Additional corrections, including flat field and a background matching at 24\,$\mu$m, were performed 
using MOPEX \citep{makovoz05}. At 70\,$\mu$m column mean subtraction and time filtering 
were applied following \citet{gordon07}.
BCD data were coadded with the MOPEX software. 
Output mosaics had pixels with sizes of 2\farcs5, 4\arcsec and 8\arcsec at 24, 70, and 160\,$\mu$m, respectively.
Modified version of the IDLPHOT routines were used to detect sources on the final maps and to perform 
aperture photometry (including aperture correction). 
All of our sources were detected at 24$\mu$m, and all but \object[HD 53842]{HD\,53842} at 70$\mu$m. Both 160\,$\mu$m observations 
provided upper limits.  
In the case of non-detections the positions derived from the 24$\mu$m images were used at 70 and 
160$\mu$m for the position of the photometry. 
Photometric results are summarized in Table~\ref{photdata}.

\paragraph{IRS:}
The observations were performed in standard staring mode using
high-accuracy blue peak-up. 
The data reduction began with intermediate {\sl droopres} products of 
the SSC pipeline (version S15.3 for the earlier four spectra; and S18.02 and S18.1 for the follow-up observations 
of HD\,169666 and HD\,13246, respectively). 
We then used the SMART reduction package \citep{higdon04}, in combination with IDL routines
developed for the FEPS Spitzer Science Legacy program \citep{meyer06}.  
The data reduction
steps (including error budget calculation) are outlined in detail in \citet{bouwman08} and have been successfully used in a
series of publications \citep[e.g. ][]{apai05,bouwman06,pascucci08,pascucci09}.

\section{Results} 

\subsection{Spectral energy distributions and infrared excesses}

We compiled the spectral energy distribution of each star
by combining IR fluxes from the IRS and MIPS observations with published optical and near-IR 
photometric data ({\sl Hipparcos, Tycho2, 2MASS} catalogs). 
In Figure~\ref{SEDs} we present the IR data, as well as stellar photospheric predictions obtained by
fitting an ATLAS9 atmosphere model \citep{castelli03} to the optical and near-IR measurements.
 Metallicity data from \citet{holmberg07} and computed surface gravity values, used as inputs to the fits, 
are listed in Table~\ref{stellarprop}. 
 Since from distance estimates based on {\sl Hipparcos} parallax data \citep{vanleeuwen07} 
all our stars lie in the Local Bubble \citep{lallement03}, we assumed the extinction to be negligible.
The fitted effective temperatures are included in Table~\ref{stellarprop}. 
Predicted photospheric fluxes for the MIPS bands are listed in Table~\ref{photdata}.
All our sources exhibit significant ($>3\sigma$) excess over the predicted photosphere at 24$\mu$m. 
At 70$\mu$m \object[HD 53842]{HD\,53842} was not detected, while the 
measured fluxes of the three other sources were found to be in excess.

The IRS spectra show that the excesses are present even at shorter wavelength as they 
begin to depart from the expected photosphere at $\sim$9, 14, 14 and 8$\mu$m in the case of 
\object[HD 13246]{HD\,13246}, \object[HD 53842]{HD\,53842}, \object[HD 152598]{HD\,152598} and \object[HD 169666]{HD\,169666}, 
respectively. 

\subsection{Spectral features in the IRS spectra}

Figure~\ref{IRS} presents the IRS spectrum of \object[HD 169666]{HD\,169666}.
Several broad spectral features at around 11.3, 16.4, 23.7, 27.5, and 33.8\,$\mu$m can be recognized. 
For comparison we plotted the spectrum of \object[HD 69830]{HD\,69830} and comet Hale-Bopp, which 
also exhibit features at similar wavelengths \citep{beichman05,crovisier97}. These peaks, especially 
the 11.3\,$\mu$m one, can be attributed to forsterite 
\citep{koike03,jager98}, 
indicating the presence of micron-sized silicate grains in the disk of \object[HD 169666]{HD\,169666}.  
The spectra of \object[HD 13246]{HD\,13246}, \object[HD 53842]{HD\,53842}, and \object[HD 152598]{HD\,152598} 
do not show any spectral features exceeding the noise. 
The lack of features in these disks may imply the depletion of small grains.

\subsection{Search for variability} \label{varisect}

We compared the two spectra of \object[HD 169666]{HD\,169666} obtained at 
different epochs. Both the continuum level and the spectral features were found to be remarkably unchanged. 
The flux differences were typically in the order of 1\%--2\% over the whole spectral range. 
A similar comparison for \object[HD 13246]{HD\,13246} showed more pronounced differences 
especially between 10 and 20$\mu$m. However, these two measurements were processed with different pipeline 
versions, namely with S15.3 and S18.1. We performed an analysis of four calibration measurements of $\eta^1$\,Dor  
processed with the same pipeline versions as HD\,13246. The results indicated that the differences 
observed at HD\,13246 could be explained by systematic calibrational effects related to the two 
pipeline versions.
Thus, we conclude that no significant spectral change can be detected in our measurements.

\subsection{Modeling the observed infrared excess}

We assume that the detected IR excess originates from optically thin dust confined into 
a circumstellar ring.
The excess above the predicted photosphere 
was fitted by a single temperature blackbody \citep[as e.g.][]{wyatt05,rhee08}. 
For the fitting process the IRS spectra were robustly averaged 
in five bins centered at 8, 13, 18, 24, and 32$\mu$m (the last bin was not used in the case of \object[HD 169666]{HD\,169666} 
because of the presence of 
a relatively strong feature). 
For \object[HD 13246]{HD\,13246} and \object[HD 169666]{HD\,169666}, where two IRS spectra are available, 
we used the earlier ones because they were obtained closer in time to the MIPS observations.
An iterative method 
was used to compute and apply color corrections for the MIPS data {\citep[see][]{moor06}. 
The derived dust temperatures ($T_{dust}$) are listed in Table~\ref{stellarprop} and 
the best fitting models are overplotted in Figure~\ref{SEDs}.

Assuming that the emitting grains act like blackbody
we estimated the 
ring radius using the following formula \citep{backman93}: 
\begin{equation}
\frac{R_{dust}}{AU} = \left(\frac{L_{star}}{L_{sun}}\right)^{0.5} \left(\frac{278\,K}{T_{dust}}\right)^2
\end{equation}
The fractional dust luminosities of the four systems were calculated as $f_d = L_{dust} / L_{bol}$ and are in the range of $0.3-1.8\times10^{-4}$.
The derived disk properties are listed in Table~\ref{stellarprop}. Note, that using the second epoch spectra for 
\object[HD 13246]{HD\,13246} and \object[HD 169666]{HD\,169666} the fitted disk parameters are not altered significantly.

\section{Discussion}

\subsection{Comparison with steady state evolutionary models}

In all four systems the inferred radii of dust rings correspond to the region of the asteroid belt and the gas giants in 
our solar system. A comparison of the PR and the collision timescales for our dust rings 
\citep[Equations~(14) and (15) in][]{backman93} implies that the grain evolution is dominated by collisions. The derived short lifetimes 
with respect to the ages of the stars demonstrate the second generation ('debris') nature of the dust grains, and indicates 
the presence of planetesimals co-located with the dust rings.    
\citet{wyatt07} used an analytical steady state model to study the collisional evolution of planetesimal belts and argue that
at any given age there is a maximum disk mass and fractional luminosity, since initially more massive disks consume their mass faster.
We computed the maximum fractional luminosity value ($f_{max}$) for each 
system taking into account its properties 
as listed in Table~\ref{stellarprop} and using 
Equation~(20) in \citet{wyatt07}, also adopting their fixed model parameters (belt width: $dr/r=0.5$, planetesimal strength: 
$Q^*_D$\,=\,200\,J\,kg$^{-1}$, 
planetesimal eccentricity: $e=0.05$, diameter of largest planetesimal in cascade: $D_c=2000$\,km).
Comparing these values with the calculated fractional luminosities we find that the disk properties
of \object[HD 13246]{HD\,13246}, \object[HD 53842]{HD\,53842}, and \object[HD 152598]{HD\,152598} are 
consistent with a steady-state evolutionary scenario within the uncertainties of the model. 
In contrast, in the case of \object[HD 169666]{HD\,169666} a ratio of $f/f_{max}>100$ implies 
a high dust content around this old star, inconsistently with a steady-state model.

\subsection{Origin of debris dust in individual systems}

\noindent 
{\it HD\,13246 and HD\,53842} 
belong to the Tuc-Hor association, implying an approximate age of 30\,Myr.  
In such young stars the disks are thought to be the sites of 
intense planet formation.  
In regions where the growing size of the largest planetesimals reach $\sim$2000\,km these protoplanets stir their surroundings 
leading to catastrophic collisions among the left-over planetesimals. A collisional cascade begins and 
produces large amounts of debris dust. Since the formation of giant planetesimals takes longer at larger radii 
the site of debris production propagate outwards with time \citep{kb02,kb04}. 

Using the evolutionary model of \citet{kb04} we calculated the radius of such a ring for an age of 30\,Myr, 
 as a function of initial disk surface density. We found that the 3.5--5\,AU radii derived for our disks 
can be explained by disk surface density of 
$\Sigma<0.3\Sigma_{MMSN}$, where $\Sigma_{MMSN}$ 
is the surface density of the minimum solar mass nebula. This parameter implies a relatively low initial disk mass provided that the 
disk is really in this evolutionary phase.  
After this peak the average dust production drops, 
but there may be episodic spikes due to individual collisions between left-over rocky protoplanets. 
If the dust rings around \object[HD 13246]{HD\,13246} and \object[HD 53842]{HD\,53842} represent 
these later phases of the evolution, the initial surface density could have been higher.

\medskip
\noindent 
{\it HD\,152598}
is older than the previous two stars, and at its age of ~0.21Gyr most
likely represent a later evolutionary phase, when dust grains are already
generated by colliding minor bodies in a steady planetesimal ring.

\medskip
\noindent
{\it HD\,169666}
is a new member of a distinguished, small group of FGK-type stars harboring warm debris disks of 
unusually high fractional luminosity. 
Only a handful of such systems have been identified so far: BD+20\,307, HD\,23514, HD\,69830, HD\,72905 
and $\eta$\,Corvi \citep{song05,rhee08,beichman05,beichman06,wyatt05}. 
The peculiar brightness of these disks is usually explained by an episodic increase in dust production \citep{wyatt07}. 
It is also a common property of \object[HD 169666]{HD\,169666} and most other members of this group that they are old 
($>$300\,Myr) and 
their mid-IR spectra exhibit spectral features attributed to small silicate dust grains.
Such small grains move and evolve primarily under the influence of radiation pressure and collisions.

In order to draw a picture on the dust evolution we computed the maximum size of dust grains blown out of 
the \object[HD 169666]{HD\,169666} system due to the radiation pressure. Approximating the absorption+scattering 
cross section by the geometric cross section and adopting a density of 2.5\,g\,cm$^{-3}$  
and an albedo of 0.1, we found a 'blowout-size' of 1.6\,$\mu$m \citep{burns79}.
Under the influence of radiation pressure alone these small grains would be removed very fast:
first-order estimates of the grain removal from the mid-IR emitting region (T$>130$ K) suggest typical timescales of a few months
to a year. With the most prominent opacity source removed so rapidly, significant changes in the mid--IR spectral features are to be
expected on similarly  short timescales. 
However, the mid--IR spectrum of \object[HD 169666]{HD\,169666} showed no significant variations 
in the course of three years. The constant flux levels of the spectral features indicate that
the small grains are continuously replenished on timescales of a few years or less. The existence of 
warm -- small or large -- dust in this system covers an even longer interval. 
Convolving our IRS spectrum with 25\,$\mu$m filter profile of IRAS \citep{beichman88} and comparing this synthetic 
flux value of 112\,mJy with the IRAS measurement in 1983 \citep[101$\pm$13\,mJy,][]{moshir89}, 
we could not see any variation within the measurement 
uncertainties. This finding suggests that the transient phase of \object[HD 169666]{HD\,169666} lasts 
at least for a quarter of a century. It is an open question whether the small grain production is active 
also on this longer timescale. Nevertheless, the fact that most old transient systems (see above) exhibit 
spectral features might suggest that small grains are present, and their production mechanism is active 
over a significant fraction of the transient phase. 
Because no multi-epoch measurements exist for the other five warm disks, possible variations of the spectral features 
cannot be excluded. Details of the variability depend on the actual mechanism of small grain replenishment 
in the system.

Three possible mechanisms have been put forward to explain the origin of dust in systems like \object[HD 169666]{HD\,169666}.
First, breakup of a large planetesimal similar to the event leading to the birth of the Veritas asteroid family \citep{nesvorny03}.
The probability of such an event, however, is relatively low at an age of $\sim$2\,Gyr \citep{wyatt07}.
Second, dust may originate from the continuous evaporation of a super-comet as proposed for the case of 
\object[HD 69830]{HD\,69830} \citep{beichman05}. The dust temperature in the disk of \object[HD 169666]{HD\,169666} exceeds 110\,K, 
the sublimation temperature of comets, 
consistently with this scenario. The third mechanism would be that the observed high fractional luminosity of these disks may be the result of an 
event similar to the Late Heavy Bombardment \citep{wyatt07}. 
Modeling the expected brightness and spectral evolution of transient disks around older stars in the 
three scenarios and comparing them 
with multi-epoch observations may help to identify the processes responsible for replenishing the observed massive 
dust belts in these extrasolar planetary systems.

\acknowledgments
We are grateful to G. Bryden for sending us the spectrum of HD\,69830.
We thank the anonymous referee for her/his useful suggestions.
This work is based on observations made with the {\sl Spitzer Space
Telescope}, which is operated by the Jet Propulsion Laboratory, California Institute
of Technology under a contract with NASA. Support for this work was provided by
NASA through contract 1311495 to Eureka Scientific. 
I.P. and D.A. acknowledge support through the {\sl Spitzer} NASA/RSA contract number 1351891. 
The research of \'A.K. is supported by the Netherlands Organization
for Scientific Research. Financial support from the Hungarian OTKA grants K62304 and T49082 is acknowledged.

{\it Facilities:} \facility{Spitzer}.

\clearpage

\begin{figure*} 
\plotone{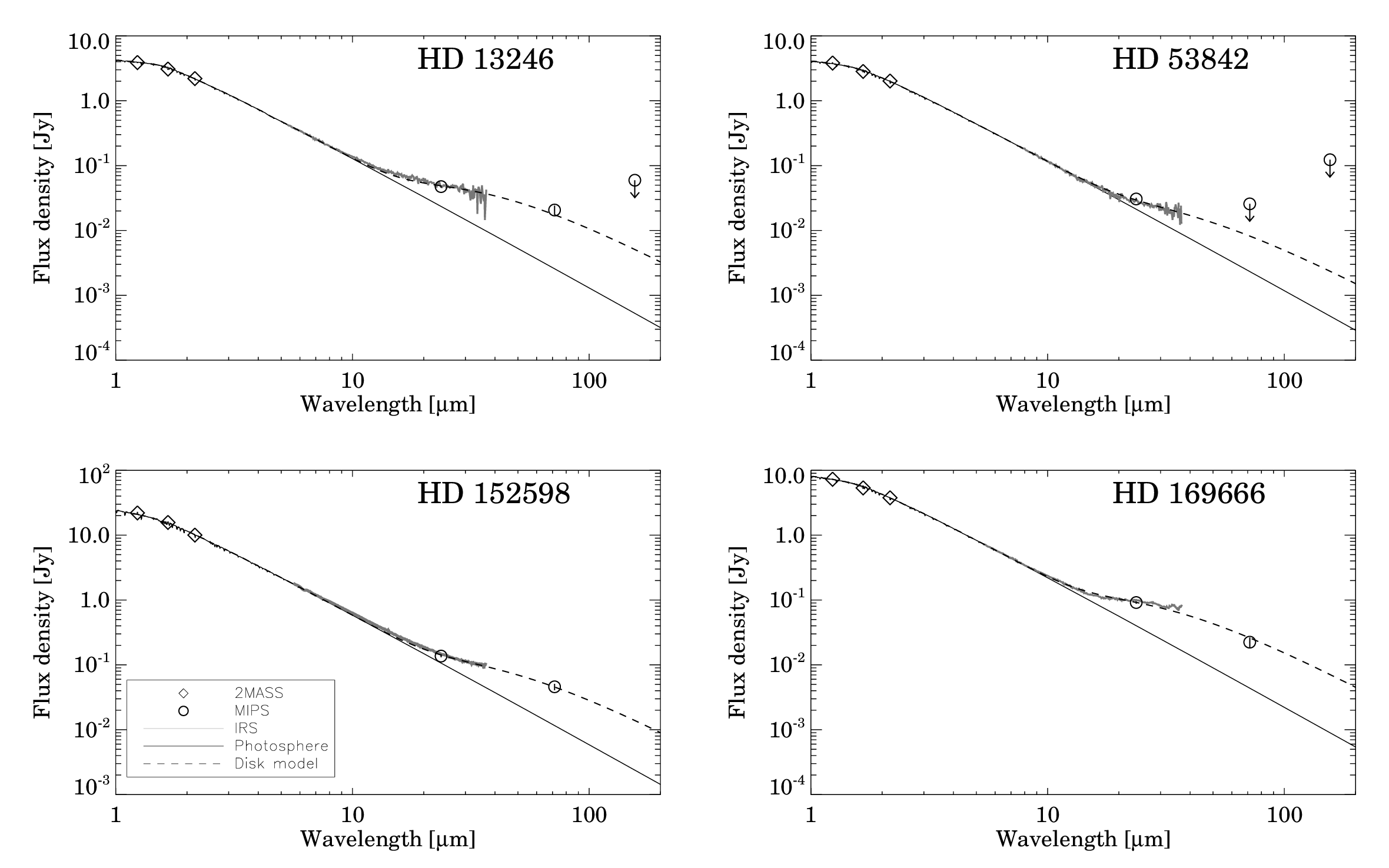}
\caption{Spectral energy distribution of the four F-type stars harboring 
warm debris disks. In the cases of HD\,13246 and HD\,169666 IRS spectrum of the first 
epochs from 2005 were plotted. 
 }
\label{SEDs}
\end{figure*}

\begin{figure} 
\plotone{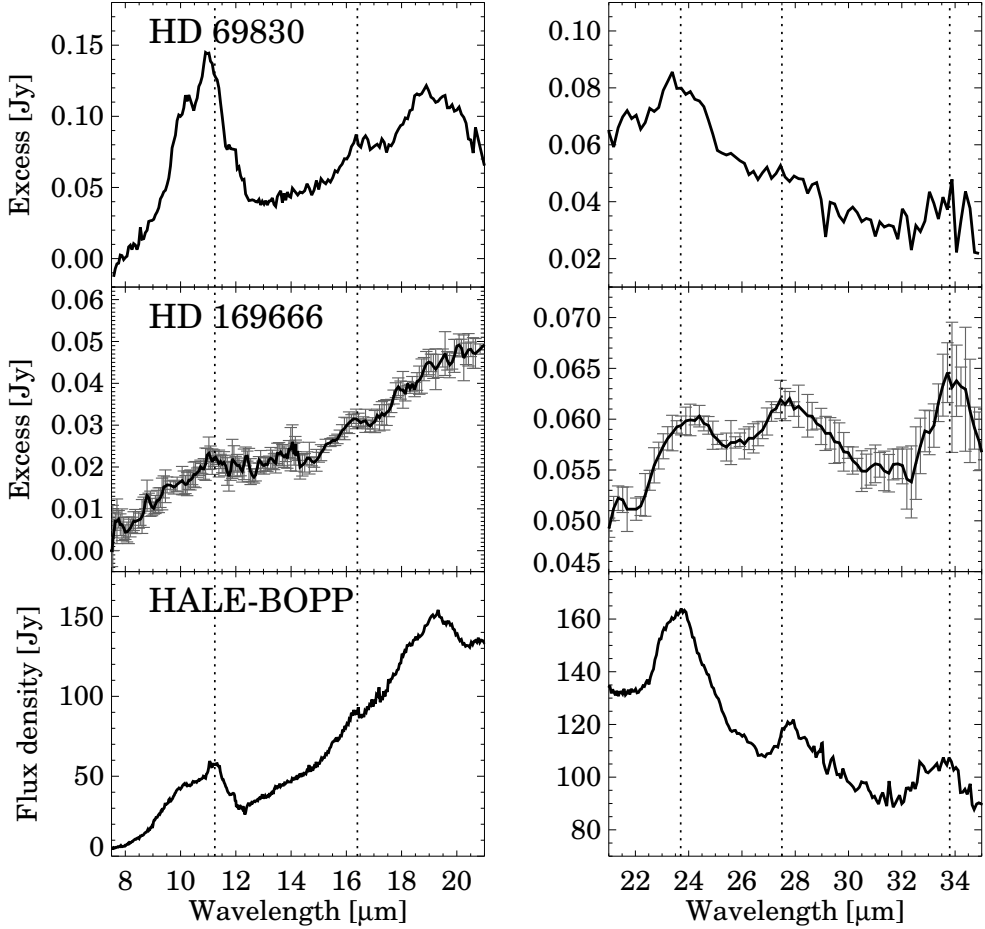}
\caption{IRS spectrum of HD\,169666 obtained on 2005 July 2. 
 The spectrum was smoothed with  
boxcar average of 5 pixels and the photospheric contribution was subtracted. 
For comparison {\sl Spitzer} spectrum of HD\,69830 \citep{beichman05} and the ISO spectrum of comet Hale-Bopp 
\citep{crovisier97} are displayed. 
The vertical dotted lines mark the laboratory peak wavelengths of 
forsterite \citep{koike03}.  \label{IRS}}
\end{figure}

\clearpage

\begin{table}
\begin{center}
\scriptsize
\caption{Stellar and disk properties \label{stellarprop}}
\begin{tabular}{lcccccccccccccc}
\tableline\tableline
          & \multicolumn{9}{c}{STELLAR PROPERTIES} & & \multicolumn{4}{c}{DISK PROPERTIES} \\
\cline{2-10}
\cline{12-15}
Source ID &     V   & SpT &  D  &${\rm T_{eff}}$ & $\rm L_*$  & $\rm M_*$ & $\rm \log{g}$ & [Fe/H] & Age &         & $\rm T_{dust}$ & $\rm R_{dust}$ & $\rm f$ & $\rm f_{max}$\\
          & (mag)   &     & (pc)&  (K)           &   ($\rm L_\sun$)& ($\rm M_\sun$) & (cm\,s$^{-1}$) & (dex)  & (Myr) &        &    [K]         &     [AU]       &  [$10^{-4}$] &  [$10^{-4}$] \\
\tableline
HD\,13246\tablenotemark{a}  &  7.50 & F8V & 44.2 & 6140  & 1.55  &  1.06 & 4.4  &  -0.17 & $\sim$30 &    &   166$\pm$18	 & 3.5$\pm$0.9         &  1.7  & 0.8 \\
HD\,53842                   &  7.46 & F5V & 56.0 & 6430  & 2.6   &  1.20 & 4.3  &  -0.16 & $\sim$30 &    &   151$\pm$24	& 5.4$\pm$1.4	      &  0.53	& 1.5 \\
HD\,152598\tablenotemark{b} &  5.34 & F0V & 29.2 & 7100  & 4.8   &  1.43 & 4.3  &  -0.21 & 210$\pm$70 &  & 135$\pm$11   & 9.3$\pm$1.5		&  0.35  &  0.5 \\
HD\,169666                  &  6.68 & F5  & 53.2 & 6540  & 4.7   &  1.35 & 4.1  &  -0.04 & $\rm 2100^{+200}_{-400}$ & & 198$\pm$13  & 4.2$\pm$0.6  &  1.84   &  0.008 \\
\tableline
\end{tabular}  
\tablenotetext{a}{HD\,13246 has a late type (K5) companion, CD-60 416 (HD\,13246B), 
with a separation of $\sim$52\arcsec~on the sky.}
\tablenotetext{b}{HD\,152598 is identified as a binary system in the CCDM catalog \citep{dommanget02}. 
However, according to the NOMAD catalog, 
the proper motion of the proposed companion (CCDM J16530+3142B) deviates significantly from the proper motion of 
HD\,152598, implying that a physical connection between these stars is unlikely.}
\end{center}
\end{table}

\begin{table}
\begin{center}
\scriptsize
\caption{Predicted photospheric fluxes and {\sl Spitzer}/MIPS photometry \label{photdata}}
\begin{tabular}{lccccccccc}
\tableline\tableline
Source ID  & P24 & F24 & P70 & F70 &  P160 & F160 \\
	   & \multicolumn{6}{c}{[mJy]}   \\
\tableline
HD\,13246  &  23.4 &  46.2$\pm$1.9 & 2.6  & 20.1$\pm$3.3 & 0.5 & $<$60.6 \\
HD\,53842  &  21.2 &  30.1$\pm$1.3 & 2.3  & $<$27.2  & 0.5 & $<$123.2	\\ 
HD\,152598 & 106.5 & 134.8$\pm$5.5 & 11.6 & 44.3$\pm$5.0 & ...   &  ...   \\ 
HD\,169666 &  40.2 &  89.2$\pm$3.7 & 4.4  & 21.8$\pm$4.0 & ...   &  ...  \\ 
\tableline
\end{tabular} 
\tablecomments{Predicted (P24, P70, P160) and measured (F24, F70, F160) flux density values 
given in mJy. The final uncertainties were computed by adding quadratically the individual 
internal errors and the absolute calibration uncertainties. 
Following the MIPS Data Handbook, we 
adopted 4\% and 7\% calibration uncertainty at 24 and 70$\mu$m. For sources not detected at a specific wavelength, 
upper limits were computed as the flux measured in the source aperture + 3$\sigma$ (all measured fluxes were positive).}
\end{center}
\end{table}


\begin{thebibliography}{}
\bibitem[Apai et al.(2005)]{apai05} Apai, D., Pascucci, I., Bouwman, J., Natta, A., Henning, T., 
\& Dullemond, C.~P.\ 2005, Science, 310, 834
\bibitem[Asiain et al.(1999)]{asiain99} Asiain, R., Figueras, F., Torra, J., \& Chen, B.\ 1999, \aap, 341, 427 
\bibitem[Backman \& Paresce(1993)]{backman93} Backman, D. E., \& Paresce, F. 1993, in Protostars and Planets III, ed.
E. H. Levy \& J. I. Lunine (Tucson, AZ: Univ. Arizona Press), 1253
\bibitem[Beichman et al.(1988)]{beichman88} Beichman, C. A., Neugebauer, G., Habing, H. J., Clegg, P. E., \& Chester, T. J.,
eds. 1988, IRAS Explanatory Supplement (Washington, DC: GPO)
\bibitem[Beichman et al.(2005)]{beichman05} Beichman, C. A., Bryden, G., Gautier, T. N., et al. 2005, \apj, 626, 1061
\bibitem[Beichman et al.(2006)]{beichman06} Beichman, C. A., et al. 2006, \apj, 639, 1166
\bibitem[Bouwman et al.(2006)]{bouwman06} Bouwman, J., Lawson, 
W.~A., Dominik, C., Feigelson, E.~D., Henning, T., Tielens, A.~G.~G.~M., 
\& Waters, L.~B.~F.~M.\ 2006, \apjl, 653, L57 
\bibitem[Bouwman et al.(2008)]{bouwman08} Bouwman, J., et al.\ 2008, \apj, 683, 479 
\bibitem[Burns et al.(1979)]{burns79} Burns, J. A., Lamy, P. L., \& Soter, S. 1979, Icarus, 40, 1
\bibitem[Castelli \& Kurucz(2003)]{castelli03} Castelli, F., \& Kurucz, R.~L.\ 2003, Modelling of Stellar Atmospheres, 210, 20P
\bibitem[Carpenter et al.(2009)]{carpenter08} Carpenter, J.~M., et al.\ 2009, \apjs, 181, 197 
\bibitem[Chen et al.(2005)]{chen05} Chen, C. H., et al. 2005, \apj, 634, 1372
\bibitem[Crovisier et al.(1997)]{crovisier97} Crovisier, J., Leech, 
K., Bockelee-Morvan, D., Brooke, T.~Y., Hanner, M.~S., Altieri, B., Keller, 
H.~U., \& Lellouch, E.\ 1997, Science, 275, 1904 
\bibitem[Currie et al.(2007)]{currie07} Currie, T., Kenyon, 
S.~J., Rieke, G., Balog, Z., \& Bromley, B.~C.\ 2007, \apjl, 663, L105
\bibitem[Currie et al.(2008)]{currie08} Currie, T., Plavchan, P., \& Kenyon, S.~J.\ 2008, \apj, 688, 597
\bibitem[Daniel et al.(1994)]{daniel94} Daniel, S.~A., Latham, D.~W., Mathieu, R.~D., \& Twarog, B.~A.\ 1994, \pasp, 106, 281 
\bibitem[Dommanget \& Nys(2002)]{dommanget02} Dommanget, J. \& Nys, O. 2002, Observations et Travaux 54, 5
\bibitem[Giardino et al.(2008)]{giardino08} Giardino, G., Pillitteri, I., Favata, F., \& Micela, G.\ 2008, \aap, 490, 113
\bibitem[Gomes et al.(2005)]{gomes05} Gomes, R., Levison, H.~F., Tsiganis, K., \& Morbidelli, A. 2005, \nat, 435, 466
\bibitem[Gordon et al.(2007)]{gordon07} Gordon, K.~D., et al.\ 2007, \pasp, 119, 1019 
\bibitem[Gorlova et al.(2007)]{gorlova07} Gorlova, N., et al. 2007, \apj, 670, 516
\bibitem[Higdon et al.(2004)]{higdon04} Higdon, S.~J.~U., et al.\ 2004, \pasp, 116, 975 
\bibitem[Hines et al.(2006)]{hines06} Hines, D. C., et al. 2006, \apj, 638, 1070
\bibitem[Holmberg et al.(2007)]{holmberg07} Holmberg, J., Nordstr{\"o}m, B., \& Andersen, J.\ 2007, \aap, 475, 519 
\bibitem[Houck et al.(2004)]{houck04} Houck, J.~R., et al.\ 2004, \apjs, 154, 18
\bibitem[J\"ager et al.(1998)]{jager98} J{\"a}ger, C., Molster, F.~J., Dorschner, J., Henning, Th., Mutschke, H., \& Waters, L.~B.~F.~M. \ 1998, \aap, 339, 904 
\bibitem[Kenyon \& Bromley(2002)]{kb02} Kenyon, S.~J., \& Bromley, B.~C.\ 2002, \apjl, 577, L35
\bibitem[Kenyon \& Bromley(2004)]{kb04} Kenyon, S.~J., \& Bromley, B.~C.\ 2004, \apjl, 602, L133
\bibitem[Koike et al.(2003)]{koike03} Koike, C., Chihara, H., Tsuchiyama, A., Suto, H., Sogawa, H., 
\& Okuda, H.\ 2003, \aap, 399, 1101 
\bibitem[Lallement et al.(2003)]{lallement03} Lallement, R., Welsh, B.~Y., Vergely, J.~L., Crifo, F., \& Sfeir, D.\ 2003, \aap, 411, 447 
\bibitem[Lisse et al.(2008)]{lisse08} Lisse, C., Chen, C., Wyatt, M. C., Morlock, A. 2008, \apj, 673, 1106
\bibitem[Makovoz \& Marleau(2005)]{makovoz05} Makovoz, D., \& Marleau, F. 2005, \pasp, 117, 1113 
\bibitem[Meyer et al.(2006)]{meyer06} Meyer, M.~R., et al.\ 2006, \pasp, 118, 1690
\bibitem[Meyer et al.(2008)]{meyer08} Meyer, M.~R., et al.\ 2008, \apjl, 673, L181 
\bibitem[Mo{\'o}r et al.(2006)]{moor06} Mo{\'o}r, A., 
{\'A}brah{\'a}m, P., Derekas, A., Kiss, C., Kiss, L.~L., Apai, D., Grady, 
C., \& Henning, T.\ 2006, \apj, 644, 525 
\bibitem[Moshir et al.(1989)]{moshir89} Moshir, M., et al. 1989, Explanatory Supplement to the IRAS Faint Source
Survey (Pasadena, CA: JPL)
\bibitem[Moultaka et al.(2004)]{moultaka04} Moultaka, J., 
Ilovaisky, S.~A., Prugniel, P., \& Soubiran, C.\ 2004, \pasp, 116, 693 
\bibitem[Nesvorn{\'y} et al.(2003)]{nesvorny03} Nesvorn{\'y}, D., 
Bottke, W.~F., Levison, H.~F., \& Dones, L.\ 2003, \apj, 591, 486
\bibitem[Pascucci et al.(2008)]{pascucci08} Pascucci, I., Apai, 
D., Hardegree-Ullman, E.~E., Kim, J.~S., Meyer, M.~R., 
\& Bouwman, J.\ 2008, \apj, 673, 477 
\bibitem[Pascucci et al.(2009)]{pascucci09} Pascucci, I., Apai, 
D., Luhman, K., Henning, T., Bouwman, J., Meyer, M.~R., Lahuis, F., 
\& Natta, A.\ 2009, \apj, 696, 143
\bibitem[Rhee et al.(2008)]{rhee08} Rhee, J. H., Song, I., \& Zuckerman, B. 2008, \apj, 675, 777
\bibitem[Rieke et al.(2004)]{rieke04} Rieke, G.~H., et al.\ 2004, \apjs, 154, 25
\bibitem[Smith et al.(2008)]{smith08} Smith, R., Wyatt, M. C., \& Dent, W. R. F. 2008, \aap, 485, 897
\bibitem[Song et al.,(2005)]{song05} Song, I., Zuckerman, B., Weinberger, A. J., \& Becklin, E. E. 2005, \nat, 436, 363
\bibitem[van Leeuwen(2007)]{vanleeuwen07} van Leeuwen, F.\ 2007, \aap, 474, 653 
\bibitem[Wyatt et al.(2005)]{wyatt05} Wyatt, M.~C., Greaves, 
J.~S., Dent, W.~R.~F., \& Coulson, I.~M.\ 2005, \apj, 620, 492 
\bibitem[Wyatt et al.(2007)]{wyatt07} Wyatt, M.~C., Smith, R., 
Greaves, J.~S., Beichman, C.~A., Bryden, G., 
\& Lisse, C.~M.\ 2007, \apj, 658, 569 
\bibitem[Wyatt(2008)]{wyatt08rev} Wyatt, M. C. 2008, \araa, 46, 339
\bibitem[Zuckerman \& Song(2004)]{zucksong04} Zuckerman, B., \& Song, I. 2004, ARA\&A, 42, 685
\end{thebibliography}
\end{document}